# Atomistic Mechanism Underlying the Si(111)-(7×7) Surface Reconstruction Revealed by Artificial Neural-network Potential

Lin Hu[1], Bing Huang[1,*] & Feng Liu[2,*]

[1]Beijing Computational Science Research Center, Beijing 100193, China

[2]Department of Materials Science and Engineering, University of Utah, Salt Lake City, UT 84112, USA

The 7×7 reconstruction of the Si(111) surface represents arguably the most fascinating surface reconstruction so far observed in nature. Yet, the atomistic mechanism underpinning its formation remains unclear after it was discovered sixty years ago. Experimentally, it is observed *post priori* so that analysis of its formation mechanism can only be carried out in analogy with archaeology. Theoretically, density-functional-theory (DFT) correctly predicts the Si(111)-(7×7) ground state but is impractical to simulate its formation process; while empirical potentials failed to produce it as the ground state. Developing an artificial neural-network potential of DFT quality, we carried out accurate large-scale simulations to unravel the formation of the Si(111)-(7×7) surface. We reveal a possible step-mediated atom-pop rate-limiting process that triggers massive non-conserved atomic rearrangements, most remarkably, a critical process of *collective* vacancy diffusion that mediates a sequence of *selective* dimer, corner-hole, stacking fault and dimer-line pattern formation, to fulfill the 7×7 reconstruction. Our findings may not only solve the long-standing mystery of this famous surface reconstruction but also illustrate the power of machine learning in studying complex structures.

*Correspondence: B.H. (bing.huang@csrc.ac.cn) or F.L. (fliu@eng.utah.edu).

The study of solid surfaces is of paramount fundamental and practical importance, as a variety of devices involves directly the properties of film/substrate surfaces, where "the interface is the device" [1]. Therefore, it is essential to understand how surface structures, at the level of monolayers of atoms, arise and evolve and affect many surface-based exotic properties. Typically, crystalline surfaces relax and/or reconstruct into a structure differing from bulk, but usually an understandable simple structure. One exception is the famous Si(111)-(7×7) reconstruction [2], which is considered one of the most intriguing surface structures in nature, representing a rare case of 2D complex system [3,4]. As the ground state of Si(111) surface, the 7×7 reconstruction is called "dimer-adatom-stacking fault" (DAS) reconstruction [5] to signify the three types of defects in the reconstructed surface [See Fig. 1(a)]. Despite the huge amount of attention devoted to this surface [2,5-18], however, its formation mechanism has remained largely a mystery for more than half a century.

The challenge to unravel the mystery of Si(111)-(7×7) reconstruction is twofold. Experimentally, it is observed *post priori* [10-12] so that analysis of its formation mechanism can only be carried out in analogy with archaeology. Theoretically, all the current simulation methods have not been both *accurate and efficient* enough to model its formation processes. The extremely large size and high complexity prohibit a full length- and time-scale *ab initio* calculation, while all the known classical potentials failed to produce it as the ground state [19] [see Fig. 1(b)]. For these reasons, the Si(111)-(7×7) reconstruction has served as a touchstone system to benchmark the advancement of newly developed experimental tools [2,5-15] and computational methods [16-18] in characterizing its properties.

Recently, machine learning has brought us a powerful tool to solve unprecedentedly complex structures and problems [20,21]. In this Letter, we demonstrate the development of a high-dimensional artificial neural-network (ANN) potential [22,23] that accurately reproduces the *ab initio* potential-energy-surface in a wide range of complex Si structures, including, most notably, the Si(111)-(7×7) ground-state structure and energies [see Fig. 1(b) and Methods] [24]. Empowered by the accuracy of this ANN potential, coupled with the efficiency of climbing image nudged elastic band (CI-NEB) [25] and molecular dynamics (MD) methods, we are able to significantly enlarge the length scale (~17,000 atoms) and extend the time scale (about several nanoseconds) of simulations to carry out an in-depth atomic-scale mechanistic study of Si(111)-(7×7) surface. Most remarkably, we reveal a possible step-mediated atom-pop rate-limiting process that triggers *massive non-conserved* atomic rearrangements, via an intriguing process of *collective* vacancy diffusion to facilitate a sequence of *selective* dimer, corner-hole, stacking fault and dimer-line pattern formation, all featured in the 7×7 reconstruction.

The bulk-terminated Si(111)-(1×1) surface is metastable with one dangling bond per surface atom, inducing a large compressive surface stress, i.e., -0.125 eV/Å$^2$ (-0.090 eV/Å$^2$) in the surface (sub-surface) layer [26]. To relieve such stress, the 1×1 surface reconstructs into a stable surface under tension, and the ground-state 7×7 reconstruction is mainly attributed to an optimal balance between surface stress and charge transfer [16-18], adopting very complex atomic rearrangements, *i.e.*, the DAS model incorporating the following key features: (i) dimers in the third layer, (ii) adatoms on surface, and (iii) stacking fault between the second and third layer in half of the unit cell, forming a faulted half unit cell (FHUC) and unfaulted HUC (UHUC) [5,18], as shown in Fig. 1(a), with several hundreds of atoms per unit cell. Other DAS-type 3×3, 5×5 and 9×9 reconstructions have also been observed under non-equilibrium conditions [12].

We note that it is impossible to directly simulate, on-the-fly, the (1×1)-to-(7×7) reconstruction even with the efficient ANN potential we developed, due to both the length- and time-scale limitations. Therefore, one hopes to piece the whole picture together step by step, by drawing from the experimental information combined with extensive search of atomistic processes of the lowest activation energy barriers (AEBs). It is well established the 7×7 surface has an atomic density ~6% lower than the 1×1 surface. For the reconstruction to occur, it is a prerequisite that the (1×1) surface finds a way to lose atoms. Thus, we first set to find the lowest-AEB pathways for an atom to leave the (1×1) surface. We started from the clean surface, and found that the AEB for such processes is ~1.50-2.27 eV (Fig. S1 [24]), after searching different atom-pop positions and pathways. This range of AEB agrees well with DFT calculations (Fig. S1 [24]), indicating a process unlikely to occur. It is also consistent with the experiments showing that the 7×7 reconstruction usually appears and expands around surface steps [6,27] and *thermodynamically* steps can act as both source and drain for the "additional" Si atoms required for the conversion between the 1×1 and 7×7 phases [6].

So, next we turned our attention to the roles that steps may play in assisting the atoms *kinetically* to pop out of the surface. We found that when steps are present, the AEBs for atoms popping out of surface is significantly reduced, and suggest that the step-mediated atom-pop process represents the rate-limiting step (Process I in Fig. 1(c)) to trigger the 7×7 reconstruction. The popped-out atoms may leave behind a good amount of vacancies in both the upper and lower terraces. Then, a critical process of *collective* vacancy diffusion (Process II) away from the steps is found to mediate a sequence of events of dimer, corner-hole, stacking-fault and dimer-line formation, as schematically illustrated in Fig. 1(c).

Experimentally, high-temperature (~700 °C) annealing is used to achieve large-area clean 7×7 domains [12], where the parent metastable 1×1 surface, instead of more stable 2×1 surface is seen transforming into the 7×7 surface. Mostly, the [11-2] and [-1-12] steps are observed, called *U*- and *F*-type step facets [28] respectively (Fig. S2 [24]). Accordingly, we focus on two step configurations: one separating a 1×1 lower and upper terrace and the other separating a 1×1 lower and a 7×7 upper terrace. In the former case, only the *U*- step exists, and three dominant atom-pop pathways [red, blue and green arrows in Fig. 2(a)] are identified. Two for a surface atom [labeled 1 and 2 in Fig. 2(a)] popping out of the lower terrace next to the step edge, and one [labeled 3 in Fig. 2(a)] out of the upper terrace. They all have a low AEB of ~0.9 eV [see Fig. 2(b)]. For the more complex case with the 7×7 upper terrace, another three pathways are identified [see Fig. 2(c)]. Two are similar to the case in Fig. 2(a) [atom 1′ and 2′, blue and red arrows in Fig. 2(c)] next to the straight portion of the step edge; the other one with the atom [atom 4 and orange arrows in Fig. 2(c)] popping into the "corner hole" of the 7×7 upper terrace. The AEBs are ~ 0.9 eV for atom 1′ and 2′, and ~ 0.8 eV for atom 4 [see Fig. 2(d)], respectively. Note that it is only the *U*-step in Fig. 2(c) that mediates a low-barrier atom-pop process, but not so for the *F*-step (the lowest AEB is found much higher, ~1.62 eV). This agrees with the experiment that the (7×7) reconstruction is preferred around the *U*- over the *F*-step [13].

There must be a number of atoms popping out around steps to create sufficient vacancies, so that the latter will induce a ~6% areal atomic density decrease after diffusing into the terrace. This means that the pop-out atoms can easily diffuse along and away from the step edge, which is indeed what we found. For example, the popped atoms may diffuse with a barrier of 0.71-0.76 eV along the step edge (Fig. S3 [24]) or ~1.20 eV over the step edge (Fig. S4 [24]); the latter agrees well with experiment (~ 1.14 eV) [29]. A collection of the popped atoms sticking to the step edge would lead to growth (retreat) of the step for the 1×1 and 7×7 surface conversion, as observed [27]. It will also leave behind a "vacancy row" in the lower (upper) terrace next to the step [see Fig. 1(c)]. The experimental observation that large 7×7 domains are terminated by step and/or disordered domains [12] apparently supports the existence of many vacancies.

Upon vacancy creation, an immediate consequence is that the atoms surrounding the vacancy are exposed becoming under-coordinated, which will spontaneously (i.e., without barrier) form dimers (see Fig. S5 [24]). Then, a key is to confirm that the optimal dimer density along the step edge corresponds to the required corner hole formation that defines the periodicity of the 7×7 reconstruction. So, we have determined the optimal dimer periodicity with the lowest energy to be N=3 (Fig. S6 [24]), which corresponds exactly to the periodicity of 7×7 reconstruction. A different dimer periodicity would result in a different periodicity of reconstruction, such as 5×5, under different experimental conductions. Meantime, the vacancies may diffuse towards the middle of terrace. In doing so, they must not only lower the atomic density, but also lead to formation of other key 7×7 structural features, including dimer, corner hole and stacking fault. However, although a variety of low-ABE pathways, as low as 0.6 eV (see Figs. 3-4 below), are found for vacancy diffusion, none of them can individually induce formation of the desired features. This triggered us to try out *collective vacancy diffusion* processes. Interestingly, we found that the collective vacancy diffusion away from the step may indeed be the next critical kinetic process responsible for a series of desired atom reorganization processes, including stacking fault formation in half of the area and dimer-line formation to define the boundary between the FHUC and UHUC. Such critical processes are found for all step configurations. Below, we use the case of 1×1 lower and upper terraces to elaborate in detail.

We first discuss the kinetic processes in the FHUC, as shown in Fig. 3. The initial vacancy positions are marked as black-dashed circles in Fig. 3(a), forming the first row of vacancies [VR1 in Fig. 3(a)] next to the step edge. As they diffuse into terrace, they exchange positions with the first row of surface atoms [AR1 in Fig. 3(a)] to form a new row of vacancies [VR2 in Fig. 3(b)], with an AEB of 0.62 eV. Meantime, the displaced AR1 will relax to the most stable positions (solid-cyan circles) forming a new row of AR1′ in Fig. 3(b). The AEB of the atom-vacancy exchange is the lowest we found in all possible pathways we tried. As a result, a row of alternating 8-atom ring and double-5-atom-ring structures form in the lower terrace along the step edge, as shown in Fig. 3(b). Next, the VR2 will continue to diffuse exchanging with AR2 in a similar manner [Fig. 3(b)]. With two steps of collective vacancy diffusion, some exposed third-layer atoms become under-coordinated to form dimers [D1 in Fig. 3(b)], with a energy gain of ~1.58 eV/dimer [30]. Furthermore, one can see from Fig. 3(c) that the dimerization, induced by every two steps of collective vacancy diffusion, also induces formation of a pair of alternating 8-atom-ring and double-5-atom-ring structures along a "diagonal" direction at 60º angle with the step edge that defines the boundary of 7×7 unit cell, while a periodic array of corner holes form naturally along the step edge, by reconstructing a pair of 8-atom and double-5-atom rings into one 12-atom ring.

The red-trapezoidal-frame area in Fig. 3(c) indicates what happens overall after the first two steps of vacancy diffusion. We note that two vacancies will diffuse out of this area into the UHUC, to be discussed later. Then the next two steps of vacancy diffusion will result in forming another smaller trapezoidal area [blue in Fig. 3(c)] with two more fewer vacancies. Meantime, a new pair of 8-atom ring and double-5-atom-ring bridged with a dimer form in the diagonal direction. Finally, after four times of two-step diffusion, a pair of 8-atom rings from two diagonal directions [orange arrows in Fig. 3(d)] will meet and coalesce into a 12-atom ring forming a corner hole with an AEB of ~0.27 eV. This completes the formation of FHUC. We note that the vacancy diffusion pathway identified above is not a simple exchange of atom-vacancy positions. Instead, while the vacancy [*e.g.*, A site for an *fcc* stacking in (111) direction] occupies the atom position (B site) [black arrow in Fig. 3(a)], the displaced atom moves to a new position (C site) [green arrow in Fig. 3(a)] rather than the original vacancy position (A site). So, as the vacancies sweep through, a stacking fault is left behind [Fig. 3(d)].

Now we discuss a direct exchange of vacancies and surface atoms in the UHUC, as shown in Fig. 4. To begin with, the first row of vacancies (VR1) exchange directly (green arrows) with the first row of atoms (AR1) [see Fig. 4(a)], which is found to have the lowest AEB (~0.60 eV). The exchanged atom first occupies a position next to an open twelve-atom ring structure and then pop to the step edge [blue arrow, similar to that in Fig. 2(a)] with an AEB of ~1.01 eV (Fig. S7 [24]). Then, similar exchange processes replicate with more vacancies feeding from the FHUC to the UFUC as mentioned above. The newly arrived vacancies exchange with the atoms, diffuse along the dimer line, and then pop to the step edge. Finally, after eight steps (two a group) of collective vacancy diffusion, a UHUC is formed. At the very last step, there will be a new vacancy row of 7 vacancies [VR8 in Fig. 4(b)] next to three dimers and in between two corner holes, same as the initial vacancy configuration in Fig. 3 next to step edge, to begin the next round of formation of a FHUC.

We further note that there are two different types of edge sites within one period between the two corner holes for the vacancy to start diffusion towards middle, which divide the surface into two separate diffusion areas subject to the local $C_3$ symmetry and result in two different stacking sequences as discussed. The FHUC assumes an ABB′ stacking [Fig. S8(a)] [24], and the UFUC keeps the original ABC stacking [Fig. S8(b)] [24]. The proposed vacancy diffusion processes in UFUC/FHUC is consistent with the observation of atom-vacancy pairs in experiment [11]. As the exchanged atoms pop to the step edge, they may diffuse out of the newly formed 7×7 unit-cell.

We emphasize that since the two elemental steps, *i.e.*, step-mediated atom-pop and collective vacancy diffusion, are identified with the lowest-AEB pathways, they constitute kinetically a highly feasible path towards the (7×7) reconstruction. Importantly, the physical manifestations derived from the proposed mechanism are all consistent with the available indirect experimental evidence, noting that experiment can only observe the reconstruction *post priori*. Furthermore, we have checked the final substructure formed at the end of the proposed processes, a partial (7×7) surface without adatoms, by the ANN potential compared with DFT [Fig. S11] [24], showing excellent agreement.

To conclude, by developing an artificial neural-network potential, we provide a possible mechanistic solution to the puzzle of Si(111)-(7×7) surface. We suggest experimental confirmation of our predictions by a more detailed study of the roles of "steps" and "vacancies". The series of kinetic processes we reveal is equally applicable to the Si(111)-(3×3), (5×5) and (9×9) reconstructions. Broadly, both the defect mediated lowering of energy barriers and the collective mass transport could be general kinetic mechanisms underlying the formation of other large-surface-unit-cell reconstructions as well as complex structures that involve transport and rearrangement of hundreds of atoms/molecules. We foresee the artificial-neural-network-potential approach to unravel more complex structures and processes in the future.

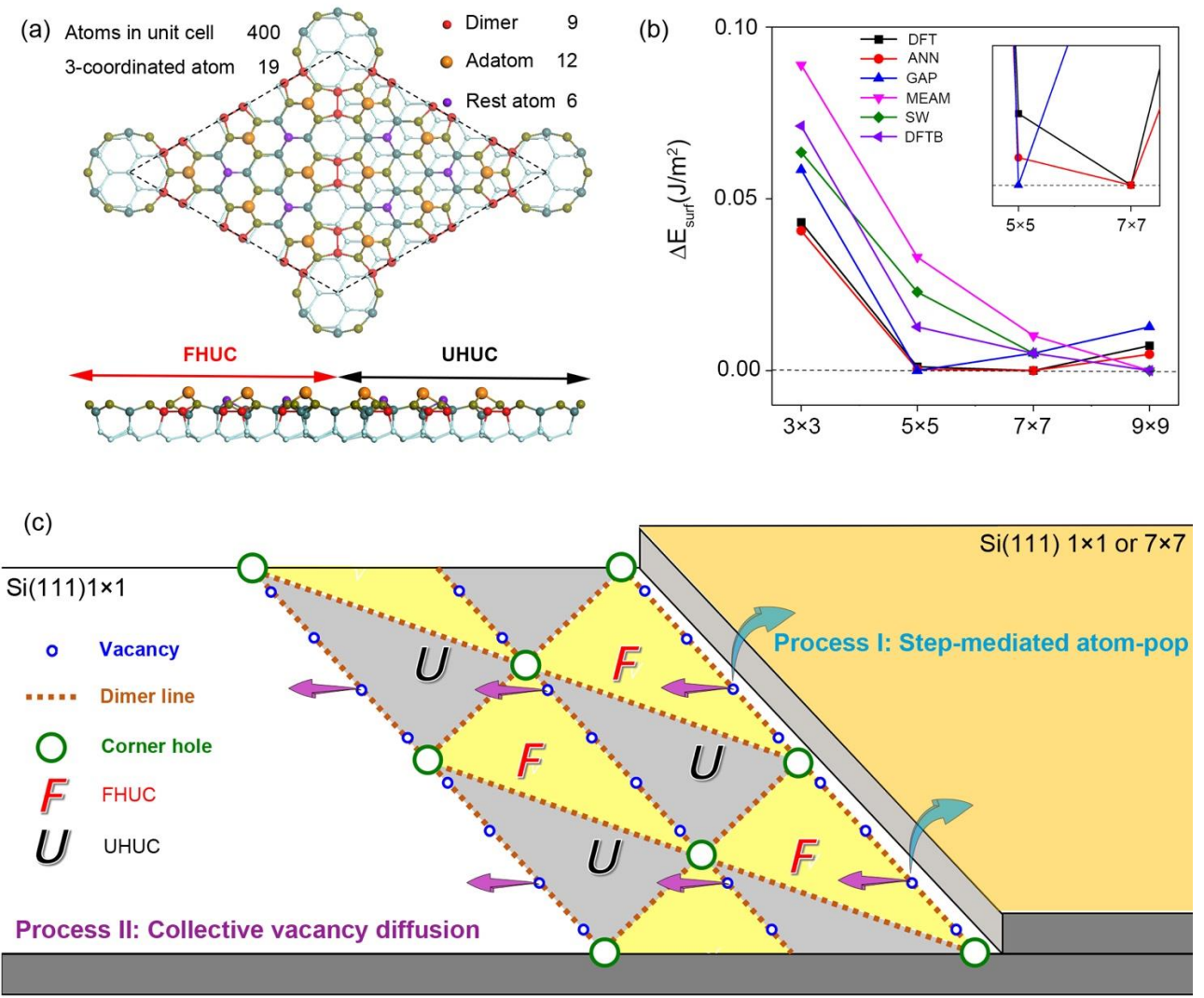

FIG. 1. (a) Top (upper panel) and side views (lower panel) of Si(111)-(7×7) surface unit cell (black-dashed line). Red, yellow and purple balls represent dimers, adatoms and rest atoms, respectively. (b) Calculated surface energies of the Si(111) surfaces with a series of DAS reconstructions, in comparison with results of DFT and other potentials, and latter are obtained from Ref. [19]. The inset is a zoom-in plot to better show the comparison between DFT, ANN and GAP potentials. (c) Illustration of step-mediated atom-pop (Process I) and collective vacancy diffusion (Process II).

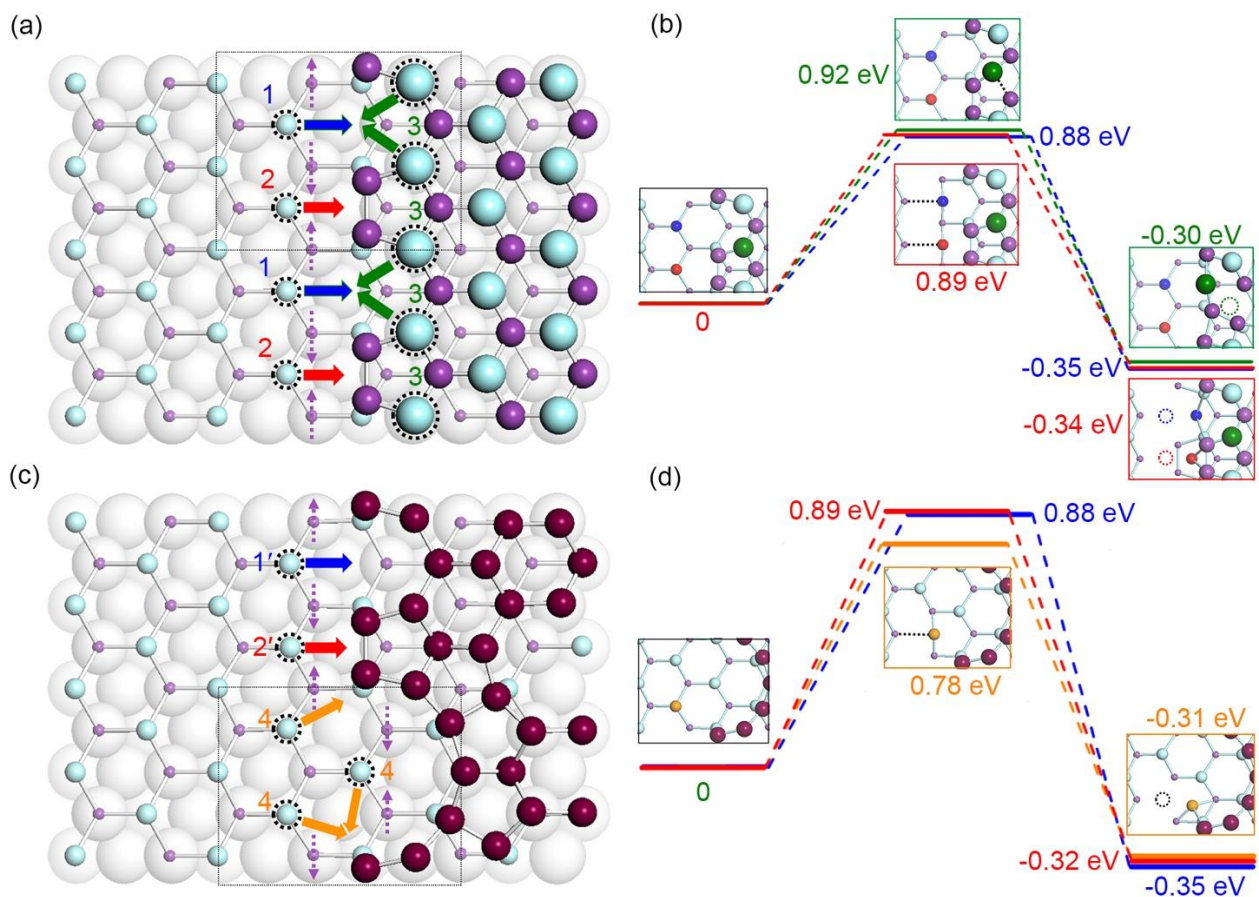

FIG. 2. (a) and (c) Atom-pop pathways around different steps. Cyan, purple and wine balls represent the second-layer, third-layer atoms of 1×1 (upper and lower) terrace and the surface-layer atoms of 7×7 upper terrace, respectively. Arrows indicate directions of atom popping or diffusing. Purple arrows indicate dimerization process. (b) and (d) Energy barriers for the kinetic processes shown by arrows in a and c, respectively, in corresponding colors. Insets show local structures of initial, transition and final states.

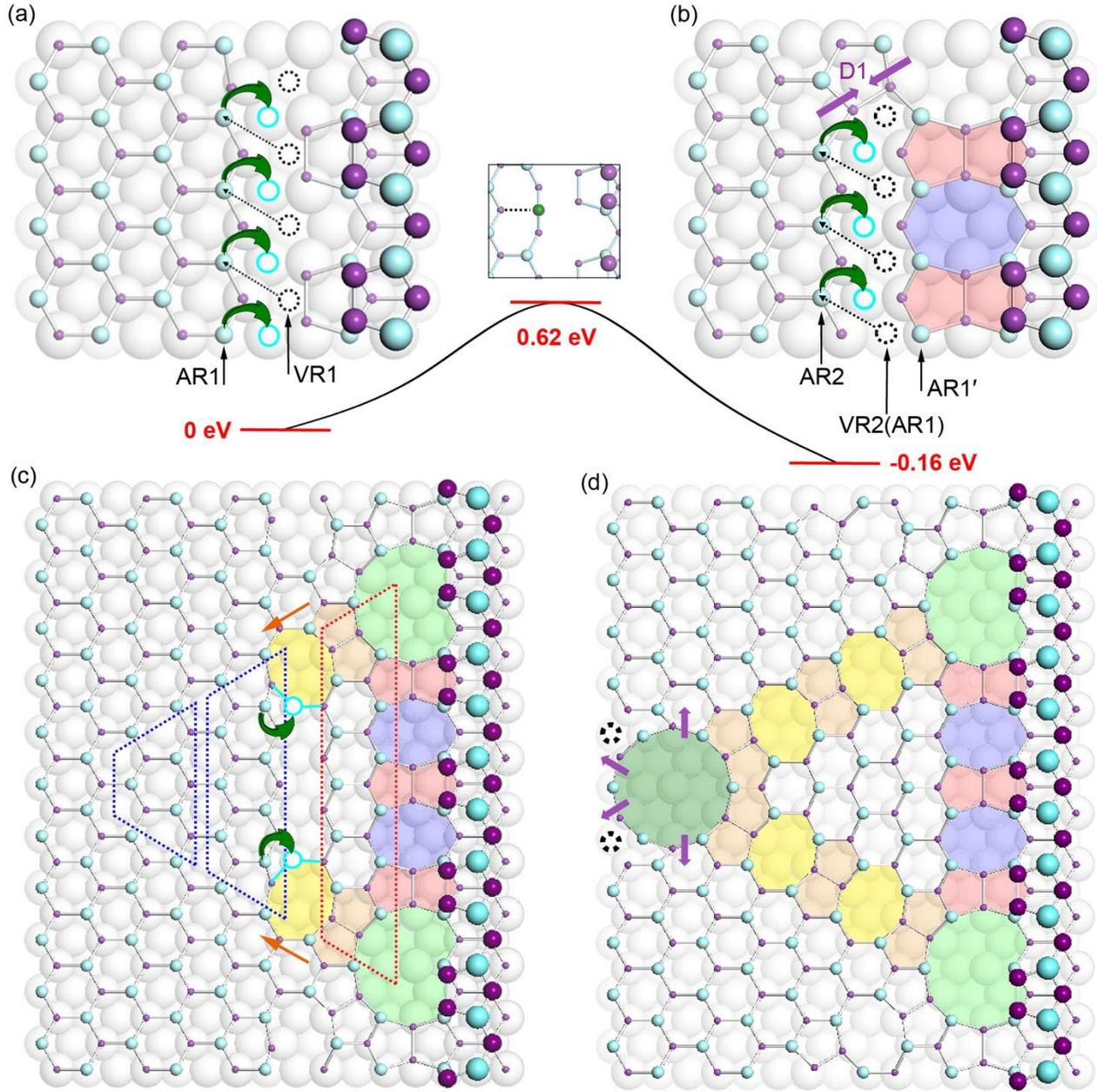

FIG. 3. (a) and (b) First and second step of vacancy diffusion process in FHUC, respectively. Black circles indicate the original vacancy positions. Black arrows indicate vacancy diffusion directions and green arrows illustrate surface atoms relaxing to new positions after exchanging with vacancies. VR1(2) and AR1(2) indicate the first (second) row of vacancies and surface atoms next to the step, respectively. Red and blue areas mark the eight-atom and double-five-atom rings, respectively. Purple arrows depict dimerization process. (c) Structure resulting from the first and second steps of *collective* vacancy diffusion. Red and blue trapezoidal frames display the areas within which the vacancies diffused in two steps. Green, brown and yellow areas mark the corner hole, eight-atom and double-five-atom rings along the diagonal directions (orange arrow), respectively. (d) Final structure of FHUC. Dark-green

area marks the twelve-atom ring forming a corner hole. Inset shows the transition state and energy barrier for vacancy diffusion indicated by arrows in (a), (b).

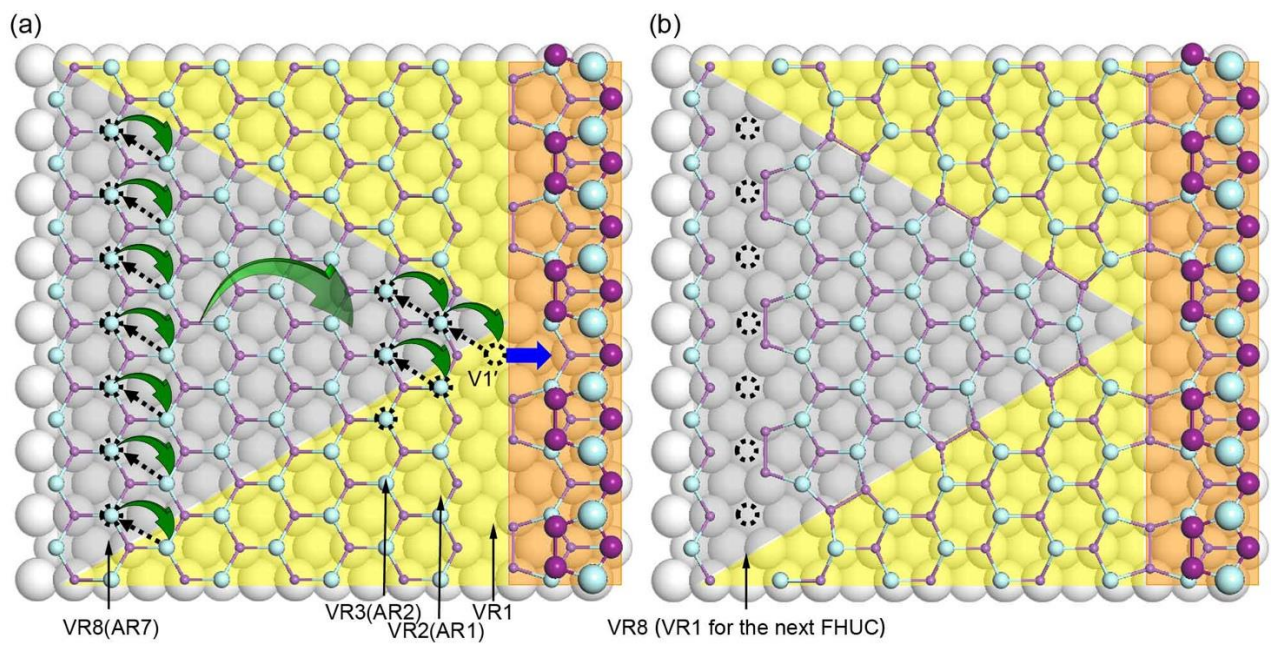

FIG. 4. (a) Vacancy diffusion process in UHUC. Gray and yellow areas mark the UHUC and FHUC, respectively. Black circles, black arrows and green arrows are same labels as Fig. 3. VR1(2, 3 and 8) and AR1(2 and 7) represent the first (second, third and eighth) row of vacancies and first (second and seventh) row of atoms next to the step, respectively. The large green arrow indicates the sequential vacancy diffusion processes leading to the VR8 formation. (b) Final structure of UHUC. VR8 becomes the "VR1" for the next round of diffusion.

**Acknowledgements** We thank S.-H. Wei, L. Kang, R. Su, B. Xu, B. Cui and N. Su for helpful discussions. L.H. and B.H. acknowledge the support from Science Challenge Project (Grant No. TZ2016003), NSFC (Grant No. 11704021) and NSAF (Grant No. U1930402). F.L. acknowledges the support from US-DOE (Grant No. DE-FG02-04ER46148). Computations were performed at Tianhe2-JK at CSRC and Paratera Clusters.